\documentclass[aps,prd,twocolumn,nobibnotes,nofootinbib]{revtex4-1} 

\expandafter\let\csname equation*\endcsname\relax 
\expandafter\let\csname endequation*\endcsname\relax 

\usepackage{amsfonts}
\usepackage{graphicx}
\usepackage{amsmath}
\usepackage{amssymb}
\usepackage{color}

\begin{document}

\title{On the Magnetosphere of an Accelerated Pulsar}

\author{T. Daniel Brennan}

\author{Samuel E. Gralla}

\address{Maryland Center for Fundamental Physics \& Joint Space-Science Institute, Department of Physics, University of Maryland, College Park, MD 20742, USA}

\begin{abstract}
We report on a remarkable class of exact solutions to force-free electrodynamics that has four-current along the light cones of an arbitrary timelike worldline in flat spacetime.  No symmetry is assumed, and the solutions are given in terms of a free function of three variables.  The field configuration should describe the outer magnetosphere of a pulsar moving on the worldline.  The power radiated is the sum of an acceleration (Larmor-type) term and a pulsar-type term.
\end{abstract}

\maketitle

\section{Introduction and Main Results}\label{sec:intro}

Pulsars are believed to be rapidly rotating neutron stars with extremely strong magnetic fields, whose pulses are caused by misalignment of the field and rotation axes.  Such a configuration is inconsistent with a vacuum exterior \cite{goldreich-julian1969}, so that pulsars must have a plasma magnetosphere.  The strong magnetic field ensures that the energy (including rest mass) and momentum of the charged particles is negligible compared to that of the fields.  Conservation then dictates that the Lorentz force density must vanish everywhere in the plasma, $F_{ab}J^b=0$.  Eliminating the current via Maxwell's equation $\nabla_b F^{ab} = 4 \pi J^a$, we may write the complete set of equations as
\begin{equation}
\nabla_{[a} F_{bc]}=0, \quad F_{ab} \nabla_c F^{bc}=0. \label{FFE}
\end{equation}
These are the equations of \textit{force-free electrodynamics}, a non-linear, deterministic  set of equations for the electromagnetic field of a magnetically dominated plasma \cite{uchida1997,komissarov2002}.

\begin{figure}[t]
\includegraphics[scale=.45]{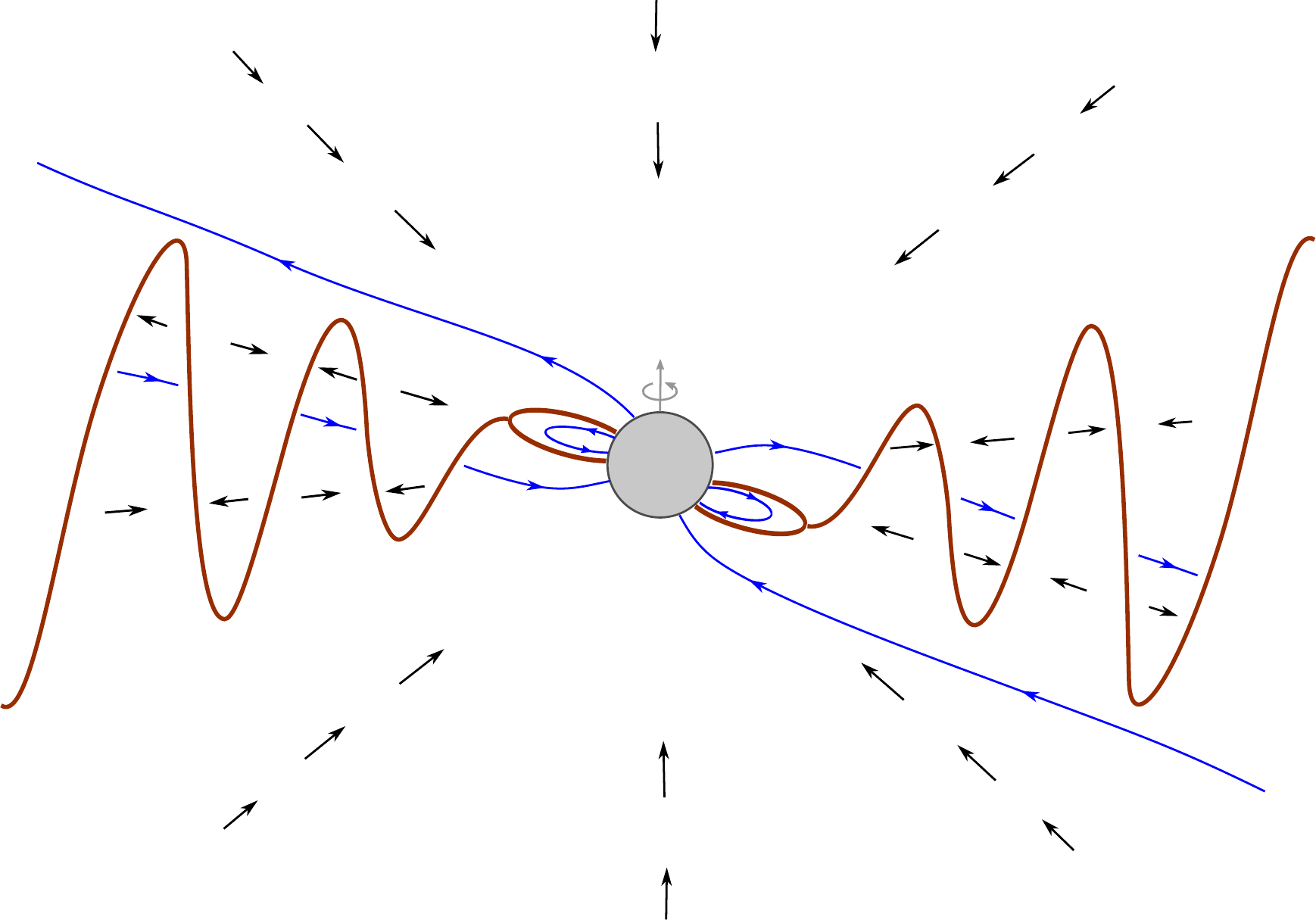}
\caption{Sketch of the pulsar magnetosphere.  Outside of a zone of closed field lines near the star, magnetic field lines (blue) run in opposite directions on opposite sides of a current sheet (brown).  (The field lines also wind around azimuthally, not shown in this projection.)  Despite the complicated geometry, the current density (black) is approximately null and radial in the open zone.}\label{fig:pic}
\vspace{-3mm}
\end{figure}

Force-free electrodynamics is very different from vacuum electrodynamics.  One dramatic example is the opening of magnetic field lines \cite{goldreich-julian1969, michel1974,ingraham1973,gralla-jacobson2014}.  If a rotating, conducting star is endowed with a magnetic dipole and immersed in vacuum, the field lines form closed loops, as usual.  If, on the other hand, the star is surrounded by a force-free plasma, lines leaving the star near its poles actually ``open up'', proceeding all the way to infinity, never to return to the star (Fig.~\ref{fig:pic}).  Closed field lines are confined to a region near the star, so that the outer magnetosphere contains only open lines, which run in opposite directions on opposite sides of a current sheet.  For aligned magnetic and rotation axes (``aligned rotor''), this sheet is on the equatorial plane, whereas for inclined axes it traces an oscillatory pattern at the rotational frequency. 

A second, key difference from vacuum electrodynamics is that stationary, axisymmetric force-free fields can transport energy and angular momentum away from an isolated source.  For example, even an aligned rotor loses energy to a force-free magnetosphere, at a rate comparable to the inclined case \cite{spitkovsky2006}.  For spinning black holes, a stationary, axisymmetric magnetosphere can extract the hole's rotational energy, as first shown by Blandford and Znajek \cite{blandford-znajek1977}.  While vacuum electrodynamics relies on acceleration to produce radiation that transports energy, force-free fields can carry away energy in steady state.

It is nevertheless natural to ask how force-free energy transport proceeds when acceleration is added to the mix.  If a magnetized neutron star is accelerated, how does the magnetosphere respond, and how is the energy output modified?  These questions have direct astrophysical application in modeling emission from compact object binaries, as potential electromagnetic counterparts to gravitational-wave observations \cite{schutz1986,holz-hughes2005} or precursor emission to gamma-ray bursts \cite{mcwilliams-levin2011}.  While numerical simulations have now successfully treated some important configurations \cite{palenzuela-etal2013,palenzuela-etal2013b,paschalidis-etienne-shapiro2013}, the high computational cost of three-dimensional runs precludes a systematic exploration using numerical techniques alone.  It is therefore of interest to develop analytical tools to address the question of the accelerated pulsar magnetosphere.

The present approach is motivated by the observation that in numerical simulations of the magnetosphere of non-moving pulsars  \cite{contopoulos-kazanas-fendt1999,spitkovsky2006,mckinney2006,timokhin2006,kalapotharakos-contopoulos-kazanas2012}, the four-current vector becomes very nearly null and radial at a few light cylinder radii from the pulsar \cite{kalapotharakos-contopoulos-kazanas2012,constantinoscomment}.  It is therefore natural to suppose that, for a pulsar in motion, the outer magnetosphere continues to host a null current pointing towards the star.  For relativistic motion, the current should point towards the pulsar location at the retarded time.  Thus we expect that the outer magnetosphere of a moving pulsar has null four-current along the light cones of the star's worldline.

In this paper we find \textit{all} solutions with null current along the light cones of a timelike worldline in flat spacetime.  We use techniques developed recently in \cite{brennan-gralla-jacobson2013}, combined with technology developed in \cite{newman-penrose1966,held-newman-posadas1970,newman1974,posadas-yanga1985}.  For fields that are smooth everywhere off the worldline, the result takes a simple form.  Let $u^a$ be the four-velocity of the worldline, extended to all spacetime by parallel transport along the (future) null cones, and let $\ell^a$ be the tangent to the null generators of the cones that satisfies $u^a \ell_a=1$.  The general solution to Eqs.~\eqref{FFE} that is smooth away from the worldline and has four-current $J^a \propto \ell^a$ is given by
\begin{equation}\label{soln}
F_{ab} = F^\textrm{q}_{ab} - 2 \ell_{[a} \nabla_{b]}\psi,
\end{equation}  
where $F^\textrm{q}_{ab}$ is the field of a magnetic monopole of charge $q$ moving on the worldline (the magnetic dual of the Lienard-Wiechert field) and $\psi$ is an \textit{arbitrary scalar field} satisfying $\ell^a \nabla_a \psi=0$.  In light of the non-linearity of Eqs.~\eqref{FFE}, it is remarkable that such a broad class of solutions can be written down analytically.  For a stationary worldline the solutions reduce to those of \cite{brennan-gralla-jacobson2013} restricted to flat spacetime, which in turn contain those of Michel \cite{michel1973b} and Lyutikov \cite{lyutikov2011} as special choices of $\psi$.  The solutions \eqref{soln} are magnetically dominated ($F^{ab}F_{ab}>0$) when $q\neq0$ and null ($F^{ab}F_{ab}=0$) when $q=0$.

The power radiated on each cone by Eq.~\eqref{soln} is
\begin{equation}\label{power}
\mathcal{P}(u) = \frac{2}{3} q^2 a^2 + \frac{1}{4\pi} \int \nabla_a \psi \nabla^a \psi \ \! dS ,
\end{equation}
where $a$ is the magnitude of the four-acceleration at the vertex of the cone (proper time $u$), and the surface of integration (area element $dS$) is a ``retarded time rest frame sphere'', the intersection of the light cone and a spacelike plane orthogonal to the four-velocity at time $u$.  (One may think of this as a sphere at future null infinity, but the integral is independent of the sphere on account of $\ell^a \nabla_a \psi=0$.)  The first term of Eq.~\eqref{power} arises from the monopole field (this is simply the Larmor formula), while the second term is due to the second term in Eq.~\eqref{soln}.  The cross term turns out to be a total derivative, and has vanished by Stokes' theorem.

Real pulsars do not contain monopoles, and the outer magnetosphere instead has a \textit{split} monopolar structure, where two regions of opposite polarity are separated by a current sheet.  The split case introduces two subtleties.  First, we may no longer assume globally smooth fields.  For fields that are only locally smooth, we find that the charge $q$ may depend on time $u$, and an additional term proportional to the time derivative $\dot{q}$ appears, Eq.~\eqref{phi2weirdo} below.  For the present work we set $\dot{q}=0$, in which case Eq.~\eqref{soln} gives the general locally smooth solution.  As explained below, $q$ corresponds to intrinsic pulsar parameters (magnetic field strength and rotation rate), so $\dot{q}=0$ restricts to pulsars whose intrinsic properties do not change significantly in time.  

The second subtlety of the split case is that Stokes' theorem fails, in general, to eliminate the cross-term in the power radiated.  In the simplest models of current sheets \cite{bogovalov1999,gralla-jacobson2014} the field strength undergoes a sign change at the sheet, so that the cross-term is continuous, and no extra terms arise.  However, in the most general case allowed by the electromagnetic junction conditions, we must supplement Eq.~\eqref{power} by boundary integrals taken on the intersection of the current sheet and the sphere, Eq.~\eqref{Psheet} below.  In \cite{kalapotharakos-contopoulos-kazanas2012} it was shown that the shape of the dipole pulsar's current sheet precisely matches the simple model of  \cite{bogovalov1999}, where the field strength undergoes a sign flip.  For this reason we expect current sheets formed in the exterior of rotating stars to generically have this simple behavior.

\begin{figure}[t]
\includegraphics[scale=.45]{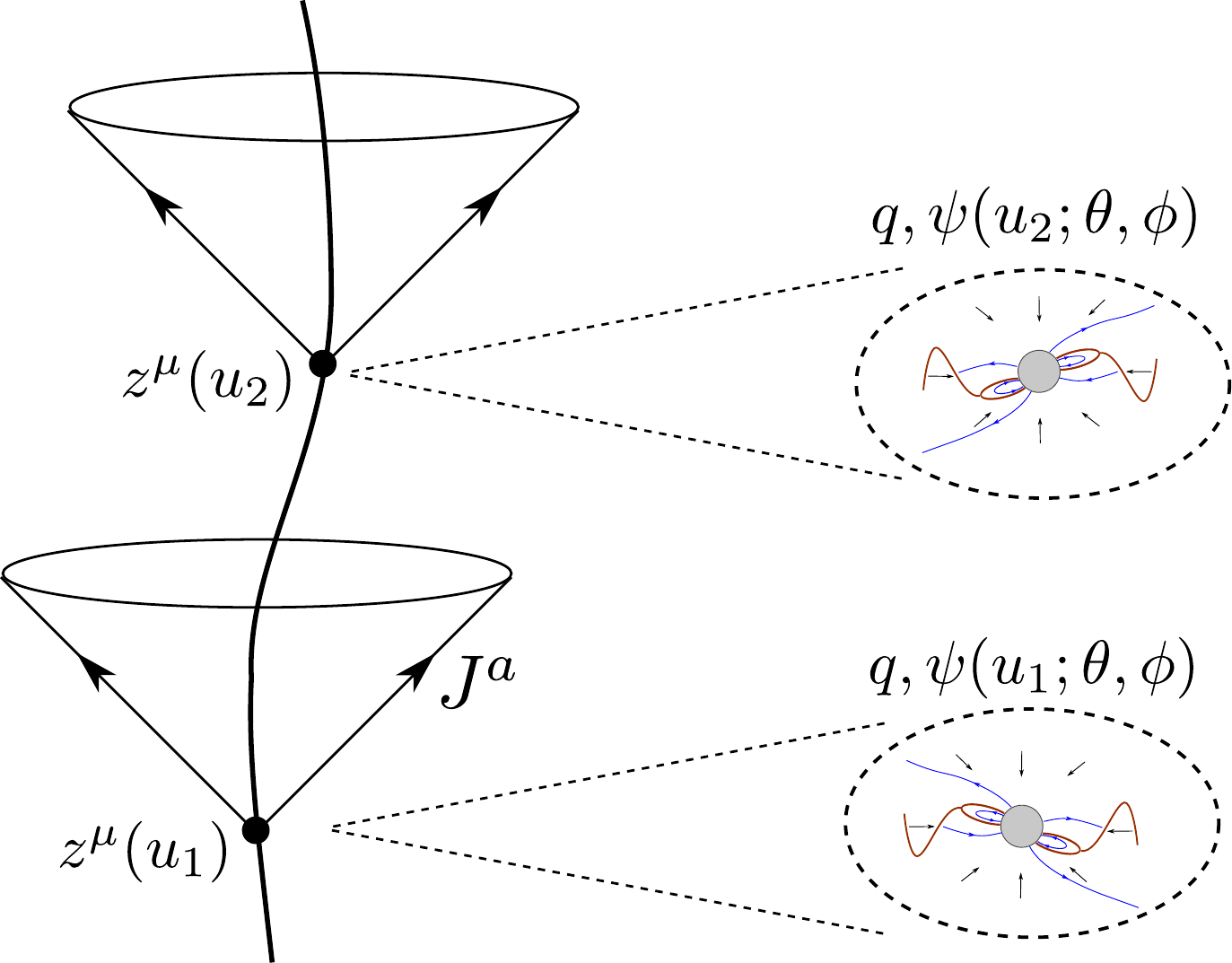}
\caption{For an accelerated pulsar we expect the far-zone four-current to be along the light cones of the worldline.  We find that exact solutions with such current are classified by a number $q$ and a function $\psi$ on the sphere cross the worldline.  These parameters therefore encode the relevant details of the near-zone pulsar physics.}\label{fig:worldline}
\vspace{-3mm}
\end{figure}

We therefore expect that the outer magnetosphere of an accelerated pulsar will be described by Eq.~\eqref{soln}, with sign reversed on either side of a current sheet,\footnote{If the pulsar is given a quadrupole or higher moment magnetic field, one would expect additional current sheets, which can also be described by our solution.} with the power radiated given by Eq.~\eqref{power}.  The solution has three free parameters/functions: the worldline, the monopole charge $q$, and the function $\psi$.  We imagine fixing these as follows.  First, perform numerical simulations of non-moving pulsars with a variety of physical parameter choices (spin, magnetic dipole, etc.) and, in each case, determine the associated $q$ and $\psi(t-r,\theta,\phi)$ by fitting the exterior magnetosphere to Eq.~\eqref{soln} with a stationary worldline.  One thus has a map between pulsar parameters at time $t$ and a function $\psi(\theta,\phi)$ on the sphere, which describes the field on the associated light cone.  Now suppose the pulsar is accelerated.  Provided the acceleration does not significantly affect the near-zone physics,\footnote{This should at least be true for small acceleration, $a \ll \Omega$, where $\Omega$ is the angular frequency of the pulsar.} one should be able to simply use the ``same'' $q$ and $\psi$ for an accelerated worldline.  That is, the same $q$ is used, while $\psi$ is promoted by demanding that it agree with the non-moving case on each light cone of the accelerated worldline (Fig.~\ref{fig:worldline}).  In this way the outer magnetosphere and radiated power can be obtained without the need to simulate the accelerated pulsar.

 


We may determine the effect of acceleration on the power radiated without performing this procedure explicitly.  The second term in Eq.~\eqref{power} agrees with the energy flux for an unaccelerated pulsar whose parameters agree instantaneously with the accelerated one.  The first term may therefore be regraded as the correction due to acceleration.  For a dipole pulsar with magnetic moment $\mu$ and angular velocity $\Omega$, dimensional analysis and linearity of the field in $\mu$ imply that $q \propto \mu \Omega$ and $\psi \propto \mu \Omega^2$.  The second term in Eq.~\eqref{power} has the usual pulsar energy loss scaling $\mu^2 \Omega^4$, while the first term gives the acceleration correction as
\begin{equation}\label{Paccel-intro}
\mathcal{P}_{\textrm{accel.}} =\frac{2}{3} q^2 a^2 \propto \mu^2 \Omega^2 a^2.
\end{equation}
For comparison, note that the power radiated by an accelerated constant dipole in vacuum scales as $\mu^2 \dot{a}^2$, where dot is a time derivative \cite{ioka-taniguchi2000}.

In Sec.~\ref{sec:astro} we estimate the size of this effect for astrophysical binaries, concluding that it is too small to be observable with present methods.  However, it is an significant fraction of the ordinary pulsar power for binaries near merger, and it would be interesting to compare with numerical simulations of binary systems.  During the inspiral it should be possible to regard each member as approximately following an accelerated trajectory in flat spacetime, and the scaling $\mu^2 \Omega^2 a^2$ should appear as part of the energy flux.  
Thus far, numerical simulations of magnetized binaries have been performed only in the irrotational case, $\Omega \approx 0$.  It would be interesting to perform simulations with non-zero values of spin in order to see if the characteristic $\mu^2 \Omega^2 a^2$ energy flux appears.  In principle this could be distinguished from other effects like unipolar induction \cite{goldreich-lynden-bell1969} by its dependence on spin and acceleration.  Alternatively one could perform a simulation of a pulsar with an unmagnetized, non-conducting companion, where there should be no unipolar induction.


From a purely theoretical standpoint, this work provides a nice coda to the story of the pulsar magnetosphere.  Perhaps the most dramatic aspect of this story is the opening of field lines, wherein the force-free plasma converts dipoles to (split) monopoles.  In a sense, our results indicate that this conversion extends to radiation, too: An accelerated pulsar radiates not as a dipole $\mu$, but rather as a monopole $q\propto\mu \Omega$.

In Sec.~\ref{sec:technology} we review some computational technology, which we use to solve the force-free equations in Sec.~\ref{sec:solution}.  We compute the energy flux in Sec.~\ref{sec:power} and discuss astrophysical applicability in Sec.~\ref{sec:astro}.  Latin indices are abstract spacetime indices (holding independent of coordinates), while Greek indices label components in a coordinate system.  The signature of our (flat) metric is $(+,-,-,-)$.

\section{Technology}\label{sec:technology}

 We begin by reviewing some technology for the light cones congruence \cite{newman-penrose1966,held-newman-posadas1970,newman1974,posadas-yanga1985}.  Consider flat spacetime in Cartesian Minkowski coordinates $x^\mu$, and let $(\zeta,\bar{\zeta})$ be complex stereographic coordinates for two-spheres in this fixed frame.  (Complex stereographic coordinates are related to spherical coordinates by $\zeta=e^{i\phi}\cot \tfrac{\theta}{2}$.)  Consider a timelike worldline parameterized by proper time $u$ as $x^\mu=z^\mu(u)$.  The four-velocity is $u^\mu=\dot{z}^\mu$, where dot denotes a $u$-derivative.  Define a new set of coordinates $(u,r,\zeta,\bar{\zeta})$ by
\begin{equation}\label{newcoords}
x^\mu=z ^\mu (u)+r \ell^\mu(u,\zeta,\bar{\zeta}),
\end{equation}
where $\ell^\mu(u,\zeta,\bar{\zeta})$ are the Minkowski coordinate components of the null vector pointing in the spatial direction $(\zeta,\bar{\zeta})$, and normalized so that $\ell_a u^a=1$.  This latter condition gives $r$ the interpretation of the spatial distance between the point $(u,r,\zeta,\bar{\zeta})$ and the worldline point $z^\mu(u)$, as measured in the rest frame of the worldline at time $u$.  Since these points are null-related, we refer to $u$ as the retarded time.  The new coordinates are defined everywhere except for the worldline $r=0$, where Eq.~\eqref{newcoords} is not differentiable.
 
 Letting an arbitrary factor $v(u,\zeta,\bar{\zeta})$ absorb the normalization, we may write $v \ell^\mu = \{1,\hat{n}(\zeta,\bar{\zeta})\}$, where $\hat{n}=\vec{x}/|\vec{x}|$ is the radial unit vector in the fixed frame. In terms of $(\zeta,\bar{\zeta})$ we then have
\begin{align}\label{lformula}
\ell^\mu=\frac{1}{vP}\left(P,\zeta+\bar{\zeta},\frac{\zeta-\bar{\zeta}}{i},\zeta \bar{\zeta}-1 \right),
\end{align}
where $P=1+\zeta \bar{\zeta}$.  If we write $u^\mu=\gamma\{1,\vec{\beta}\}$ then $\vec{\beta}$ is the three-velocity of the worldline relative to the fixed frame.  From $u^a \ell_a=1$ and $v \ell^\mu = \{1,\hat{n}\}$ we then obtain the explicit formula $v=\gamma(1-\vec{\beta}\cdot \hat{n})$.  This form helps for checking a convenient identity satisfied by $v$, 
\begin{equation}\label{videntity}
v^{-2} = 1 + \Delta \log v,
\end{equation}
where $\Delta$ is the Laplacian on the unit two-sphere.

In Eq.~\eqref{newcoords}, $r \ell^\mu$ is naturally regarded as a ``displacement vector'' between $z^\mu(u)$ and the field point $x^\mu$.  For later purposes it is convenient to let $\ell^a$ be the vector field whose Minkowski coordinates are given by Eq.~\eqref{lformula} at each point $(u,r,\zeta,\bar{\zeta})$ on the manifold.  This vector field is tangent to the congruence of future-directed null geodesics emanating from the worldline, and ill-defined on the worldline itself.  We will also regard $v(u,\zeta,\bar{\zeta})$ as a scalar field on the manifold (minus the worldline), whose particular space-time dependence encodes the three-velocity $\vec{\beta}(u)$.  Finally we extend the four-velocity off the worldline by parallel transport along the light cones, i.e., $u^\mu(u,r,\zeta,\bar{\zeta})=\dot{z}^\mu(u)$.

To compute the metric components in the new coordinates it is useful to note that $\partial_\zeta \ell^\mu$ is complex-null and orthogonal to $\ell^\mu$ and $u^\mu=\dot{z}^\mu$.  From Eq.~\eqref{newcoords} we then find
\begin{align}\label{metric}
ds^2=(1-2r\frac{\dot{v}}{v})du^2+2 du dr-\frac{4r^2}{v^2 P^2}d\zeta d\bar{\zeta}.
\end{align}
The metric of a (unit) two-sphere in complex stereographic coordinates is given by $4P^{-2} d\zeta d\bar{\zeta}$.  The extra factor of $v^2$ in Eq.~\eqref{metric} reflects the fact that the two-surface $u=r=\textrm{const}$ is a constant-distance sphere in the rest frame associated with retarded time $u$, whereas $(\zeta,\bar{\zeta})$ were defined relative to the fixed frame.  We refer to $u=r=\textrm{const}$ spheres as rest frame spheres.

By construction, we have $\ell = \partial_r$ in the new coordinates.  From Eq.~\eqref{metric} we may select three other null vectors satisfying the Newman-Penrose (NP) \cite{newman-penrose1962} requirements $\ell^a n_a=1$ and $m^a \bar{m}_a=-1$ (other inner products vanishing),
\begin{subequations}\label{tetrad}
\begin{align}
\ell^\mu &=(0,1,0,0)\\
n^\mu&=\left(1,-\frac{1}{2}\left(1-2r\frac{\dot{v}}{v}\right),0,0\right)\\
m^\mu&=\left(0,0,\frac{v P}{\sqrt{2}r},0\right)\\
\bar{m}^\mu&=\left(0,0,0,\frac{v P}{\sqrt{2}r}\right).
\end{align}
\end{subequations}
The vectors $\ell$ and $n$ are null normals to the rest frame spheres, while $m$ and $\bar{m}$ are complex-null tangents.  From Eq.~\eqref{newcoords}, the $(u,r,\zeta,\bar{\zeta})$ coordinate components of the extended four-velocity are
\begin{align}\label{xidot}
u^\mu = (1,r\frac{\dot{v}}{v},0,0) = n^\mu + \tfrac{1}{2} \ell^\mu.
\end{align}
Finally, a unit vector orthogonal to $u^a$ and to rest frame spheres is given in these coordinates by
\begin{align}\label{R}
R^\mu = (1,-1-r\frac{\dot{v}}{v},0,0) = n^\mu - \tfrac{1}{2} \ell^\mu.
\end{align}

The spin coefficients for the tetrad \eqref{tetrad} are
\begin{align}
\rho=2\mu=-\frac{1}{r}, \quad \alpha = -\bar{\beta} = \frac{\partial_{\bar{\zeta}}(P v)}{2\sqrt{2}r},\nonumber \\ \quad \gamma = -\frac{\dot{v}}{2v}, \quad \nu=-\frac{Pv}{\sqrt{2}} \partial_{\bar{\zeta}}\left( \frac{\dot{v}}{{v}} \right),  \label{spin-coefficients}
\end{align}
with all other coefficients vanishing.  From $\kappa=\sigma=\textrm{Im}[\rho]=0$ we see that the null congruence along $\ell$ is geodesic, shear-free, and twist-free \cite{newman-penrose1962}.  Using $u^a=n^a+\tfrac{1}{2} \ell^a$ and the NP equations \cite{newman-penrose1962} we may compute the frame components of the four-acceleration $a^a=u^b \nabla_b u^a$ (extended off the worldline by parallel transport along null cones), finding $a^a=\nu m^a + \bar{\nu} \bar{m}^a + \gamma(\ell^a - 2 n^a)$.  Two particularly useful quantities are the projection on to $\ell^a$,
\begin{equation}\label{aell}
a^a \ell_a = \frac{\dot{v}}{{v}},
\end{equation}
and the magnitude,
\begin{equation}\label{a2}
a^a a_a = - P^2 v^2 \partial_\zeta \left( \frac{\dot{v}}{v} \right) \partial_{\bar{\zeta}} \left( \frac{\dot{v}}{v} \right) - \left( \frac{\dot{v}}{v} \right)^2.
\end{equation}

A final bit of technology we will find useful are the $\eth$ and $\bar{\eth}$ operators.  These operators are defined on functions with a definite spin weight as
\begin{align}
\eth \eta & = P^{1-s} \partial_\zeta (P^s \eta) \label{eth} \\
\bar{\eth} \eta & = P^{1+s} \partial_{\bar{\zeta}} (P^{-s} \eta). \label{ethbar}
\end{align}
The spin-weight $s$ of a function $\eta$ is refers to its behavior $\eta\rightarrow \exp[i \theta s] \eta$ under rotations of the sphere tetrad vectors $m \rightarrow \exp[i \theta]m$.  The application of $\eth$ raises the spin-weight of a quantity by one, while $\bar{\eth}$ lowers by one.  Any smooth function of spin-weight $-1$ can be written as $\eth$ of a spin-weight zero function (and similarly for $\bar{\eth}$ and spin-weight +1).  Acting on spin-weight zero functions, we have $\eth \bar{\eth}=\bar{\eth} \eth=\Delta$, where $\Delta=P^2 \partial_\zeta \partial_{\bar{\zeta}}$ is the sphere Laplacian.  Both $\eth$ and $\bar{\eth}$ obey the Leibniz rule and have the property that the sphere-integral of $\eth f$ (or $\bar{\eth}f$) is vanishing for any spin-weighted function $f$.  Thus total derivatives may be freely thrown away under integrals.  In anticipation of this use we rewrite Eq.~\eqref{a2} as
\begin{equation}\label{larmorsavior}
\eth \left( \frac{\dot{v}}{v} \right) \bar{\eth} \left( \frac{\dot{v}}{v} \right) = - \frac{2}{3} \frac{a^a a_a}{v^2} + \frac{1}{6} \eth \bar{\eth} \left(\frac{\dot{v}}{v}\right)^2,
\end{equation}
where $v$ and $\dot{v}$ have spin-weight zero, and Eq.~\eqref{videntity} has been used.

\section{Solution}\label{sec:solution}
 We follow the general approach of \cite{brennan-gralla-jacobson2013}, using some techniques from \cite{newman1974,posadas-yanga1985}.  The electromagnetic NP scalars are defined as \cite{newman-penrose1962}
\begin{subequations}\label{NPdef}
\begin{align}
\phi_0=&F_{ab} \ell^a m^b \\
\phi_1=&\tfrac{1}{2} F_{ab}(\ell^a n^b +\bar{m}^a m^b)\\
\phi_2=&F_{ab} \bar{m}^a n^b.
\end{align}
\end{subequations}
We assume that the current is along the null congruence, $J^a = \mathcal{J} \ell^a$ with $\mathcal{J}\neq 0$.  The force-free condition $F_{ab}J^b=0$ then becomes $F_{ab} \ell^b=0$, or equivalently
\begin{equation}\label{FFphi}
\phi_0=0, \qquad \textrm{Re}[\phi_1]=0.
\end{equation}
Using Eqs.~\eqref{FFphi}, \eqref{spin-coefficients} and \eqref{tetrad} in the spin-coefficient version of Maxwell's equations \cite{teukolsky1973}, we find
\begin{gather}
\left( \partial_r + \frac{2}{r} \right) \phi_1 = 0, \quad \chi \partial_\zeta \phi_1  = 0, \label{espresso}\\
\left( \partial_r + \frac{1}{r} \right) \phi_2 - \chi \partial_{\bar{\zeta}} \phi_1  = 0, \label{americano}\\
\chi^2 \partial_\zeta\left(\frac{\phi_2}{\chi}\right) - \left( \partial_u - \frac{1}{2} \left(1-2r\frac{\dot{v}}{v} \right) \partial_r - \frac{1}{r} \right) \phi_1 = 2 \pi \mathcal{J},\label{tea}
\end{gather}
where $\chi=(v P)/(\sqrt{2} r)$.  

Eqs.~\eqref{espresso}, together with the fact that $\phi_1$ is pure imaginary, imply that $\phi_1=i q(u)/(2r^2)$ for a real function $q(u)$, where the factor of $1/2$ is convenient.  This function must be independent of angles $(\zeta,\bar{\zeta})$ to satisfy the equations locally, but our application to split monopole magnetospheres requires us to allow different local solutions to be patched together, so that $q(u)$ becomes a piecewise-constant function on the sphere.  We will write $q(u;\mathcal{D})$ to remind the reader that this function may take different constant values on different domains $\mathcal{D}$ of the sphere,
\begin{equation}\label{phi1}
\phi_1 = \frac{i q(u;\mathcal{D})}{2r^2}.
\end{equation}
The magnetic monopole charge $Q$ of a field configuration may be defined as $1/(4\pi)$ times the magnetic flux through a closed surface.  (For regular fields satisfying Maxwell's equations such an integral is always zero, but our fields are singular on the worldline.) From Eqs.~\eqref{NPdef} together with the fact that $m$ and $\bar{m}$ span rest frame spheres $\mathcal{S}$, we see that
\begin{equation}\label{qtot}
Q = \frac{1}{4\pi} \int_{\mathcal{S}} F = \int q(u;\mathcal{D}) \ d\Omega,
\end{equation}
where the first statement views $F$ as a two-form.  A solution with non-zero $Q$ cannot be realized physically, since it would require the presence of a magnetic monopole charge on the worldline.  (In the picture of matching the solution to a pulsar interior, the matching could only succeed if the pulsar contained magnetic monopoles.)  Thus for physical solutions we require that
\begin{equation}\label{nomonopole}
\int q(u;\mathcal{D}) \ d\Omega = 0.
\end{equation}
If we regard $q(u;\mathcal{D})$ as the effective local monopole charge, this statement means that the total effective monopole charge must vanish.  

Note that we work locally on each domain $\mathcal{D}$, so that $\eth q=0$.  Requiring Eqs.~\eqref{espresso}-\eqref{tea} to also be satisfied at the domain  boundaries (i.e., in a distributional sense on spacetime) would enforce a strictly null, radial, current even on any current sheets.  This is too restrictive for the application to pulsar magnetospheres.  Instead, one must allow (non-force-free) charge-current to flow in the sheets.  We defer the specific selection of appropriate domains to the task of constructing a detailed model of an outer magnetosphere.

Plugging Eq.~\eqref{phi1} into Eq.~\eqref{americano}, we find that 
\begin{equation}\label{phi2f}
\phi_2=\frac{f(u,\zeta,\bar{\zeta})}{r}
\end{equation}
 for a complex function $f(u,\zeta,\bar{\zeta})$.  Eq.~\eqref{tea} then yields
\begin{equation}\label{cake}
\frac{1}{\sqrt{2}}\eth\left(\frac{f}{v}\right) - \frac{iq}{2}\partial_u\left(\frac{1}{v^2}\right) = 2 \pi \frac{r^2}{v^{2}} \mathcal{J} + \frac{i \dot{q}}{2 v^2},
\end{equation}
where we use the $\eth$ operator introduced in Eq.~\eqref{eth}, and $f/v$ has spin-weight $-1$.  Using the identity \eqref{videntity} we may express Eq.~\eqref{cake} as
\begin{equation}\label{torte}
\eth\left(\frac{f}{\sqrt{2}v} - \frac{iq}{2} \bar{\eth} \frac{\dot{v}}{v} \right) = 2 \pi \frac{r^2}{v^{2}} \mathcal{J} + \frac{i \dot{q}}{2 v^2}.
\end{equation}
Since the term in parentheses is a spin $-1$ function of $(u,\zeta,\bar{\zeta})$, it may be expressed as $\tfrac{1}{2} \bar{\eth} \psi$ for a spin-zero function $\psi(u,\zeta,\bar{\zeta})$ (with a convenient factor of $\tfrac{1}{2}$), 
\begin{equation}\label{cupcake}
\bar{\eth} \psi = \sqrt{2}\frac{f}{v} - iq \ \bar{\eth} \left(\frac{\dot{v}}{v}\right).
\end{equation}
Then Eq.~\eqref{torte} becomes
\begin{equation}\label{pie}
\Delta \psi = 4 \pi \frac{r^2}{v^{2}} \mathcal{J} + \frac{i \dot{q}}{2 v^2},
\end{equation}
where $\eth \bar{\eth}=\Delta$ is the sphere Laplacian.

To solve Eq.~\eqref{pie} we use the identity \eqref{videntity} and split $\psi$ into real and imaginary parts $\psi=\psi^R+i\psi^I$, finding
\begin{equation}\label{psiR}
\Delta \psi^R = 4\pi \frac{r^2}{v^2} \mathcal{J}
\end{equation}
and
\begin{equation}\label{psiI}
\Delta (\psi^I - \dot{q} \log v) = \dot{q}.
\end{equation}
Eq.~\eqref{psiR} provides no constraint on $\psi^R$, since we allow the current $\mathcal{J}$ to take on whatever value is set by the solution of Eqs.~\eqref{FFE}.  On the other hand, Eq.~\eqref{psiI} should be solved for $\psi^I$.  This equation asks for a function on the sphere whose Laplacian is constant.  There is no such globally regular solution, but using $\Delta=P^2 \partial_\zeta \partial_{\bar{\zeta}}$ it is easy to see that the general solution is
\begin{equation}\label{weirdo}
\psi^I - \dot{q} \log v = \dot{q} \log P + \alpha(\zeta) + \beta(\bar{\zeta})
\end{equation}
where $\alpha$ and $\beta$ are free functions representing the freedom of adding  homogeneous solutions.  However, since only $\bar{\eth} \psi$ appears in the field strength (see Eqs.~\eqref{cupcake} and \eqref{phi2f}), we may set $\alpha=0$ without loss of generality.  But the constraint that $\psi^I$ be real then implies that $\beta$ is constant (recall $\zeta=e^{i \phi} \cot\tfrac{\theta}{2})$, in which case we may also set $\beta=0$.  In this case combining Eqs.~\eqref{phi2f}, \eqref{cupcake}, and \eqref{weirdo} yields
\begin{equation}\label{phi2weirdo}
\phi_2 = \frac{1}{\sqrt{2}} \frac{v}{r} \bar{\eth}\left( \psi^R + iq \frac{\dot{v}}{v} + i \dot{q} \log(P v) \right).
\end{equation}

While the last term in Eq.~\eqref{phi2weirdo} diverges at the ``south pole'' $\zeta,\bar{\zeta} \rightarrow \infty$, the choice of this pole is arbitrary (picked out here by working in a particular set of coordinates), and a regular solution could be constructed by taking regular portions of this term on each domain $\mathcal{D}$.  We will discuss this type of construction in a future paper.  For this paper we set $\dot{q}=0$.

Setting $\dot{q}=0$  and collecting everything together, our solution for the NP scalars is
\begin{align}
\phi_0 & = 0 \label{phi0soln} \\
\phi_1 & = \frac{i q}{2r^2} \label{phi1soln} \\
\phi_2 & = \frac{1}{\sqrt{2}} \frac{v}{r} \bar{\eth}\left( \psi + iq \frac{\dot{v}}{v} \right), \label{phi2soln} 
\end{align}
where $\psi(u,\zeta,\bar{\zeta})$ is a free real function on the sphere cross time, and $q(\mathcal{D})$ is a piecewise constant function on the sphere.  The domains $\mathcal{D}$ may in general change with time $u$, but the value of the $q$ within each domain must remain constant.  While every choice of $\psi$ and $q(\mathcal{D})$ gives rise to a solution away from the domain boundaries, ensuring that the field satisfies appropriate junction conditions at the domains (i.e., that the boundaries do not host magnetic monopole sources) will restrict the choice.

When $\psi$ vanishes (or is constant), Eqs.~\eqref{phi0soln}-\eqref{phi2soln} are the spin coefficient form of the point monopole field \cite{newman1974,posadas-yanga1985}, locally in each domain $\mathcal{D}$.  To see that the $\psi$ term gives the correction listed in Eq.~\eqref{soln}, contract $\ell_{[a} \nabla_{b]} \psi$ with tetrad vectors to compute the associated NP scalars, using the formulae \eqref{tetrad} for our tetrad.

For a field with $\phi_0=0$, the quadratic invariants are given by $*F^{ab}F_{ab}=-4\textrm{Im}[\phi_1^2]$ and $F^{ab}F_{ab}=-8\textrm{Re}[\phi_1^2]$ \cite{brennan-gralla-jacobson2013}.  Thus our solutions have $*F^{ab}F_{ab}=0$ (which is true of any non-vacuum solution of Eqs.~\eqref{FFE}) and  $F^{ab}F_{ab}=2q^2/r^4$.  In particular the solutions are magnetically dominated ($F^{ab}F_{ab}>0$) when the monopole charge is non-vanishing, and otherwise null ($F^{ab}F_{ab}=0$).

The charge-current for the solution is given by Eq.~\eqref{pie} with $\dot{q}=0$,
\begin{equation}\label{Jsoln}
\mathcal{J} = \frac{1}{4\pi} \frac{v^2}{r^2} \Delta \psi.
\end{equation}
Integrating Eq.~\eqref{Jsoln} with respect to the sphere element $d\Omega$ shows that $\int \mathcal{J} v^{-2}d\Omega=0$, or equivalently $\int J^a dS_a=0$, where $dS_{\mu}=r^2/v^2 R_\mu d\Omega$ is the oriented area element on rest frame spheres.  Thus no three-current flows through any such sphere, and, since $J^a$ is null, the net charge on the sphere vanishes.  In particular, the worldline does not act as a source or sink of current, and the magnetosphere is charge-neutral overall.

\section{Power}\label{sec:power}

We now compute the power radiated by Eqs.~\eqref{phi0soln}-\eqref{phi2soln}, making the additional assumption that the magnitude of $q(\mathcal{D})$ is the same in each domain.  Since power is Lorentz-invariant, we may use the frame defined by the four-velocity at each retarded time.  The flux through a large rest frame sphere (sphere at future null infinity) is 
\begin{equation}
\mathcal{P}(u) = \lim_{r \rightarrow \infty} \int T_{ab} u^a R^b \frac{r^2}{v^2} d\Omega = \frac{1}{2\pi} \lim_{r \rightarrow \infty} \int |\phi_2|^2 \frac{r^2}{v^2} d\Omega.\nonumber
\end{equation}
For the second equality we have used the NP form of the electromagnetic stress-tensor $T_{ab}$ \cite{teukolsky1973}, the formulas $u^a = n^a + \tfrac{1}{2}\ell^a$ and $R^a=n^a-\tfrac{1}{2} \ell^a$, and the facts that $\phi_0=0$ and $\phi_1=O(1/r^2)$ for our solution.  Using the explicit formula \eqref{phi2soln} yields
\begin{align}
\mathcal{P}(u) & = \frac{1}{4\pi} \int \eth \left( \psi - i q \frac{\dot{v}}{v} \right) \bar{\eth} \left( \psi + i q \frac{\dot{v}}{v} \right) d\Omega \nonumber \\
& = \frac{1}{4\pi} \int \Bigg\{ \eth \psi \ \bar{\eth} \psi + q^2 \eth \left( \frac{\dot{v}}{v}\right) \bar{\eth} \left( \frac{\dot{v}}{v}\right) \nonumber \\
& \qquad \quad + i q \left[ \bar{\eth} \left( \frac{\dot{v}}{v} \eth \psi \right) - \eth \left( \frac{\dot{v}}{v} \bar{\eth} \psi \right) \right] \Bigg\} d\Omega \label{oliphaunt}
\end{align}
where we have used the Leibniz rule and the fact that $\eth$ and $\bar{\eth}$ commute on spin-zero functions.

  We call the first term in Eq.~\eqref{oliphaunt} the pulsar power.  To rewrite this term covariantly, note that the non-zero components of the inverse metric are $g^{ur}=1$, $g^{rr}=-1+2r(\dot{v}/v)$, and $g^{\zeta \bar{\zeta}}=-(P^2 v^2)/(2 r^2)$.  Furthermore, the area element on the surface of integration is $dS=-r^2/v^2 d\Omega$.  We then have $\eth \psi \bar{\eth} \psi d\Omega=P^2 \partial_\zeta \psi \partial_{\bar{\zeta}} \psi (-v^2/r^2) dS = g^{\mu \nu} \nabla_\mu \psi \nabla_\nu \psi dS$.  This latter expression is covariant, and we have
\begin{equation}\label{Ppulsar}
\mathcal{P}_{\textrm{pulsar}}(u) = \frac{1}{4\pi} \int \nabla_a \psi \nabla^a \psi \ \! dS,
\end{equation}
deriving the second term in Eq.~\eqref{power}.  While the integral arose on a large sphere, it is in fact independent of the sphere radius.

We call the second term in Eq.~\eqref{oliphaunt} the acceleration power.  To evaluate this term we use Eq.~\eqref{larmorsavior}.  Since $q^2$ is assumed constant over the sphere, $q^2 \tfrac{1}{6} \eth \bar{\eth} (\dot{v}/v)$ is a total derivative and does not contribute.  We are then left with
\begin{equation}\label{Paccel}
\mathcal{P}_{\textrm{accel.}}(u) = \frac{1}{4\pi} \int \left(- \frac{2}{3} q^2 \frac{a^a a_a}{v^2} \right) \ d\Omega = - \frac{2}{3} q^2 a^a a_a,
\end{equation}
where we pull the constants $q^2$ and $a_a a^a$ out of the integral and then use Eq.~\eqref{videntity}, throwing away the total derivative.  This derives the first term in Eq.~\eqref{power}.

We call the remaining contribution to Eq.~\eqref{oliphaunt} the sheet power.  Since $\bar{\eth}(A \eth B)-\eth(A \bar{\eth}B)$ is equal to $P^2 [ \partial_{\bar{\zeta}}(A \partial_\zeta B) - \partial_\zeta(A \partial_{\bar{\zeta}} B)]$ for spin-zero functions $A$ and $B$, we identify this contribution (last line of Eq.~\eqref{oliphaunt}) as the integral of the two-form $-2 q d(\tfrac{\dot{v}}{v}d\psi)$, where $d$ is the exterior derivative.  Then by Stokes theorem on each domain and $\dot{v}/v=a^a \ell_a$ (Eq.~\eqref{aell}), we have
\begin{equation}\label{Psheet}
\mathcal{P}_{\textrm{sheet}}(u) = \frac{-1}{2\pi} \int_S d\left( q \frac{\dot{v}}{v} d\psi \right) = \frac{-1}{2\pi}\int_{C_{\pm}} q \ \! a^a \ell_a d \psi.
\end{equation}
Here $C$ represents the oriented curve(s) on the sphere present at the boundary between domains $\mathcal{D}$.  The notation $C_{\pm}$ indicates that an integral is to be performed using the limiting value of the integrand from either side of the curve, with opposite orientations on opposite sides, as required by Stokes' theorem.  

The integral arises for a large sphere, but since neither the domains $\mathcal{D}$ nor the integrand $q a^a \ell_a d\psi$ depend on $r$, the integral does not depend on the radius of the sphere.  The curve $C$ may be characterized in an invariant manner as the intersection between the current sheet, the light cone at time $u$, and a spacelike plane orthogonal to the four-velocity.  The choice of spacelike plane corresponds to the radius $r$, and the above properties ensure that the integral is independent of this choice.  As discussed in the introduction, we expect the contributions from $C_+$ and $C_-$ to cancel for the current sheets that arise in pulsar magnetospheres, so that this term makes no contribution to the energy flux.

\section{Astrophysical Applicability}\label{sec:astro}

Our exact solutions involve the idealization that the force-free plasma fills all of space.  In reality, the force-free magnetosphere of a compact object will only extend a finite distance.  (This distance is hard to estimate, since the force-free description does not include information on the particle density.)  To apply our solutions and their predicted scaling \eqref{Paccel-intro}, the force-free magnetosphere should extend for at least several characteristic lengths of the trajectory.  In the case of a comparable mass binary, this characteristic length is the orbital radius.  

For a pulsar member of a comparable mass Newtonian binary in circular orbit, Eq.~\eqref{Paccel-intro} becomes
\begin{equation}
\mathcal{P}_{\textrm{accel.}} \approx 10^{36} \frac{B_{12}{}^2 M_{1.4}{}^2 R_{10}{}^6}{P^2D_{10}{}^4} \textrm{erg/s},
\end{equation}
where $B_{12} \times 10^{12} \textrm{ Gauss}$ is the surface magnetic field strength, $R_{10} \times 10 \textrm{ km}$ is the stellar radius, $M_{1.4} \times 1.4 \ M_{\odot}$ is the stellar mass, $P \times 1 \textrm{  s}$ is the rotational period, and $D_{10} \times 10 \textrm{ km}$ is the orbital separation.  The relative strength of the acceleration power $\mu^2 \Omega^2 a^2$ to the pulsar power $\mu^2 \Omega^4$ is
\begin{equation}\label{Pratio}
\frac{\mathcal{P}_{\textrm{accel.}}}{\mathcal{P}_{\textrm{pulsar}}} \sim \frac{a^2}{\Omega^2} \approx 10^6 \frac{M_{1.4}{}^2P^2}{D_{10}{}^4}  \approx 10^{-5} \left( \frac{M_{1.4} P^3}{P_{\textrm{orb.}}^4}\right)^{2/3},
\end{equation}
where $P_{\textrm{orb.}}$ is the orbital period of the binary in seconds. 

The energy lost due to the acceleration will come at the expense of some combination of the rotational (spin) and translational (orbital) kinetic energy of the body.  Any orbital energy decrease will be undetectably small, since $\mathcal{P}_{\textrm{accel.}}$ is vastly subdominant (by a factor of $\sim10^{-30} D_{10}$) to the power in gravitational-wave emission.  The effect on spin-down is also small, but may become relevant for binaries near merger: The ratio $\mathcal{P}_{\textrm{accel.}}/\mathcal{P}_{\textrm{pulsar}}$ can range from $\sim 10^{-15}$ for known binary pulsars ($P_{\textrm{orb.}}\sim \textrm{hours}$) all the way to order unity for binaries near merger ($P_{\textrm{orb.}}\sim .01 \textrm{ s}$).  Unfortunately, there is little prospect for receiving electromagnetic signals from this pre-merger inspiral period.

\acknowledgements{We thank Ted Jacobson, Constantinos Kalapotharakos and Maura McLaughlin for helpful conversations.  S.G. acknowledges support from NASA through the Einstein Fellowship Program, Grant PF1-120082. D.B. was supported in part by the NSF under grant No. PHY-0903572.}

\bibliography{accelpulsar}{}

\end{document}